\documentclass[amsfonts, amssymb, amsmath,twocolumn]{revtex4-1}

\usepackage{graphicx}
\usepackage{dcolumn}
\usepackage{bm}
\usepackage{color}
\usepackage[utf8]{inputenc}
\usepackage{soul}
\usepackage[T1]{fontenc}
\usepackage[ruled,vlined]{algorithm2e}
\usepackage{comment}
\usepackage[toc,page]{appendix}

\usepackage[bottom]{footmisc}

\usepackage{xparse}
\usepackage{tikz}
\usetikzlibrary{matrix,backgrounds}
\pgfdeclarelayer{myback}
\pgfsetlayers{myback,background,main}

\usepackage[colorinlistoftodos]{todonotes}
\usepackage[colorlinks=true, allcolors=blue]{hyperref}
\definecolor{capri}{rgb}{0.0, 0.75, 1.0}

\begin{document}

\preprint{APS/123-QED}

\title{Extracting real social interactions from social media: \\  a debate of COVID--19 policies in Mexico}

\author{Alberto Garc\'ia-Rodr\'iguez$^{1,2}$ \href{mailto:alberto.g.rodgz@gmail.com}{alberto.g.rodgz@gmail.com}}
\author{Tzipe Govezensky$^3$ \href{mailto:tzipe@unam.mx}{tzipe@unam.mx}}
\author{Carlos Gershenson$^{2,4,5}$ \href{mailto:cgg@unam.mx}{cgg@unam.mx}}
\author{Gerardo G. Naumis$^1$ \href{mailto:naumis@fisica.unam.mx }{naumis@fisica.unam.mx}}
\author{Rafael A. Barrio$^1$ \href{mailto:barrio@fisica.unam.mx}{barrio@fisica.unam.mx}}

\affiliation{$^1$Departamento de Sistemas Complejos, Instituto de  F\'isica, Universidad Nacional Aut\'onoma de M\'exico (UNAM), Apdo. Postal 20-364, 01000, CDMX, M\'exico}
\affiliation{$^2$ Instituto de Investigaciones en Matem\'aticas Aplicadas y en Sistemas, Universidad Nacional Aut\'onoma de M\'exico (UNAM), 01000, CDMX, Mexico.}
   \affiliation{$^3$Instituto de Investigaciones Biom\'edicas, Universidad Nacional Aut\'onoma de M\'exico (UNAM), 04510 CDMX, Mexico,}
   \affiliation{$^4$ Centro de Ciencias de la Complejidad,
Universidad Nacional Autónoma de México, Mexico City, Mexico,}
   \affiliation{$^d$ Lakeside Labs GmbH, Lakeside Park B04, 9020 Klagenfurt am Wörthersee, Austria}

\date{\today}

\begin{abstract}
A study of the dynamical formation of networks of friends and enemies in social media, in this case Twitter, is presented. We characterise the single node properties of such networks, as the clustering coefficient and the degree, to investigate the structure of links. The results indicate that the network is made from three kinds of nodes: one with high clustering coefficient but very small degree, a second group has zero clustering coefficient with variable degree, and finally, a third group in which the clustering coefficient as a function of the degree decays  as a power law. This third group represents $\sim2\%$ of the nodes and is characteristic of dynamical networks with feedback. This part of the lattice 
seemingly represents strongly interacting friends in a real social network.

\end{abstract}

\maketitle

\section{Introduction}

In principle, social media were designed to allow its members to express opinions about different topics, make new friends and conections. An interesting feature of social media is the vast amount of data collection~\cite{Pacheco2020Coordinated}. There is considerable research that takes advantage of such possibility as it brings a way to perform quantitative analysis of social relationships. For example, this has been made with weighted links in 
mobile phone calls~\cite{Iniguez_2009}. However,  in spite of these well investigated aspects of making friends and links, there is another side of the coin when dealing with human networks: enemies and conflicts~\cite{Paluck566,Timmo2017}.

When opinions are expressed, and 
particularly about polemic topics, such as politics, sports, religion, etc., there is a tendency for opinions to polarize~\cite{Sasahara2020} (a fact soon recognized by British clubs by asking its members to avoid such dividing topics). This leads to interesting features such as the echo chamber effect~\cite{Sasahara2020} and the zealot effect~\cite{Mobilia2003,Klamser2017}. 

Also, economical and geopolitical interests, historical affinities and so on, play an important vital role in this polarization effect. Moreover, this could lead to conflict escalation and
it is not unusual to observe ``arms races'' in which  social media members appeal to astroturfing~\cite{Keller2020}, disinformation~\cite{Pierri2020}, collusive behavior~\cite{Dutta2018}, payed haters, bots, bots farms and hybrid human-bots farms~\cite{Dutta2020}.
An interesting feature of social media is the possibility of studying the time-evolving network topology, which is a factor that only recently has been  taken into account in models~\cite{Clough2014,Holme2015}.

Several efforts have been made in order to distinguish
such behaviors, yet it is very difficult to determine if an enemy in Twitter is human~\cite{Context,Dutta2020}. Artificial intelligence strategies have been implemented to perform such task~\cite{Wang2010}, but in fact, its efficiencies are limited and low. Eventually, the task requires human intervention as  it is highly context-dependent and requires to mange subtleties such as sarcasm, irony, and black humor~\cite{Context}. Others try to perform the same task by taking into account the frequency of emission or number of tweets ~\cite{Dutta2020}.

From a different perspective, in a previous paper we performed an analysis of friends and foes in several schools in Mexico City~\cite{Naumis_2015}. This allowed to get a glimpse of the main differences between social media networks and a small scale social network in which non-human intervention is absent. One of the striking results of such study was the different friend and foe networks topologies\cite{Naumis_2015}. Enemy networks tend to be much more heterogeneous resulting in a kind of public enemy effect. Also, below certain age, genders tend to be dissociated, a fact explained by the Heider balance theory~\cite{Naumis_2015}. Such effects have been confirmed in other studies~\cite{Paluck566,Timmo2017}. Moreover, three body effects are crucial for conflicts in social networks~\cite{Naumis_2007,Oldenburg_2018} and thus it is worthwhile to test whether such effects are present in social media.

In this article, we explore the question of
how social based human networks are contained within social media networks. In particular, we focus on Twitter. 

The layout of this paper is the following. In Sec.~\ref{Sec:Background} we present the network analysis performed in a one year study of the Twitter generated network topology. Then in Sec.~\ref{Sec:Analysis} we analyze the network while in Sec.~\ref{Sec:Coretweet} we discuss the retweet network topology.
Finally, we provide a section with the conclusions and perspectives of the work. Details of our methods are included in Sec.~\ref{sec:methods}.

\section{Data Mining}\label{Sec:Background}

During the COVID--19 pandemic, many political decisions were implemented aiming to mitigate the effects of the virus SARS-CoV-2, producing a strong polarization in discussions about this health emergency. In this study we used this subject as an example of strong polarization of opinions in Mexico that has been present during the last year. Main actors from both sides (pro and anti-government) were selected according to the press and to the number of tweets. We collected tweets for 3 weeks of the month of May 2020 ($6{\mathrm{th}}-21^{\mathrm{st}}$) (see details in Sec.~\ref{sec:methods}). The chosen actors related to the COVID--19 health emergency in Mexico were:

\begin{description}
\item[1. amlo, lopezobrador] L\'opez Obrador current president of Mexico (2018-2024).
\item[2. morena] current governing and majority party in Mexico.
\item[3. 4t] the ``fourth transformation'', refers to changes promoted by current government.
\item[4. gatell] Hugo López-Gatell Ramírez, current Undersecretary for Prevention and Health Promotion at the Mexican Ministry of Health, in charge of Mexico's COVID--19 strategy and response.
\item[5. FelipeCalderon] Felipe Calder\'on's Twitter account. President of Mexico from 2006 to 2012 and outspoken critic of the new government through social media.
\item[6. SSalud\_mx] Official Twitter account of Mexico's Ministry of Health.
\item[7. insabi] Institute of Health for Wellbeing. 
\end{description}

In order to investigate the network produced by these actors when tweeting about the COVID--19 health emergency, 
we chose the following keywords: 

\begin{description}
\item[1] {\bf encasa (at home) , sanitaria (sanitary), sarampion (measles), cuarentena (quarantine) , salud (health), fase3 (phase 3)} Words related to messages issued by the secretary of health and emerging issues such as measles. We consider these to be ``\emph{health labels}'' (see below).
\item[2] {\bf covid, coronavirus, corona, pulmonia, neumonia, sarscov2, respir} We consider these to be ``\emph{COVID--19 labels}'' (see below).

\end{description}

We generated our core set with all those tweets that match two criteria: any aforementioned actor 1-7 and at least one of the two types of labels. In this way, we ensure that the tweet in question talks about health and is related to Mexico. In total, we collected 2,950,080 tweets.

\begin{figure}[htb]
\includegraphics[width=.9\columnwidth]{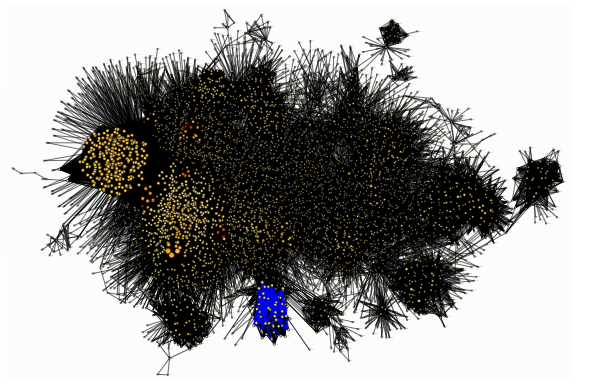}
\caption{Snapshot of a twitter discussion. Edges between nodes represent retweets or mentions to another user. We pay particular attention to the blue color subset which shows those users who connect with other users more than $8$ times. Yellow nodes have more than $5$ edges.} 
 \label{fig:fig1}
\end{figure}

In Fig.~\ref{fig:fig1} we plot the discussion network including quotes and mentions to other users. As we can see, the graph looks very dense and shapeless. Trying to 
distinguish groups within this network, we identified users who connected with other users more than 8 times in this period (in blue), and users with more than 5 edges (in yellow). However, since quotes and mentions are difficult to classify as supporting or attacking tweets, we concluded that in order to identify separate political groups, quotes and mentions should be excluded.

To obtain a directed network, we took advantage of the fact that retweets (Rt) contain information about the source and the target user, making it possible to generate directional links from user to user. In the context of retweets, the in-degree ($k_{in}$), can be interpreted as the popularity of the user, and the out-degree ($k_{out}$), as reflecting the support that a user gives to others. Groups in this network are formed by homophily so that different political ideologies will segregate.

\begin{figure}[ht!]
 \includegraphics[width=.9\linewidth]{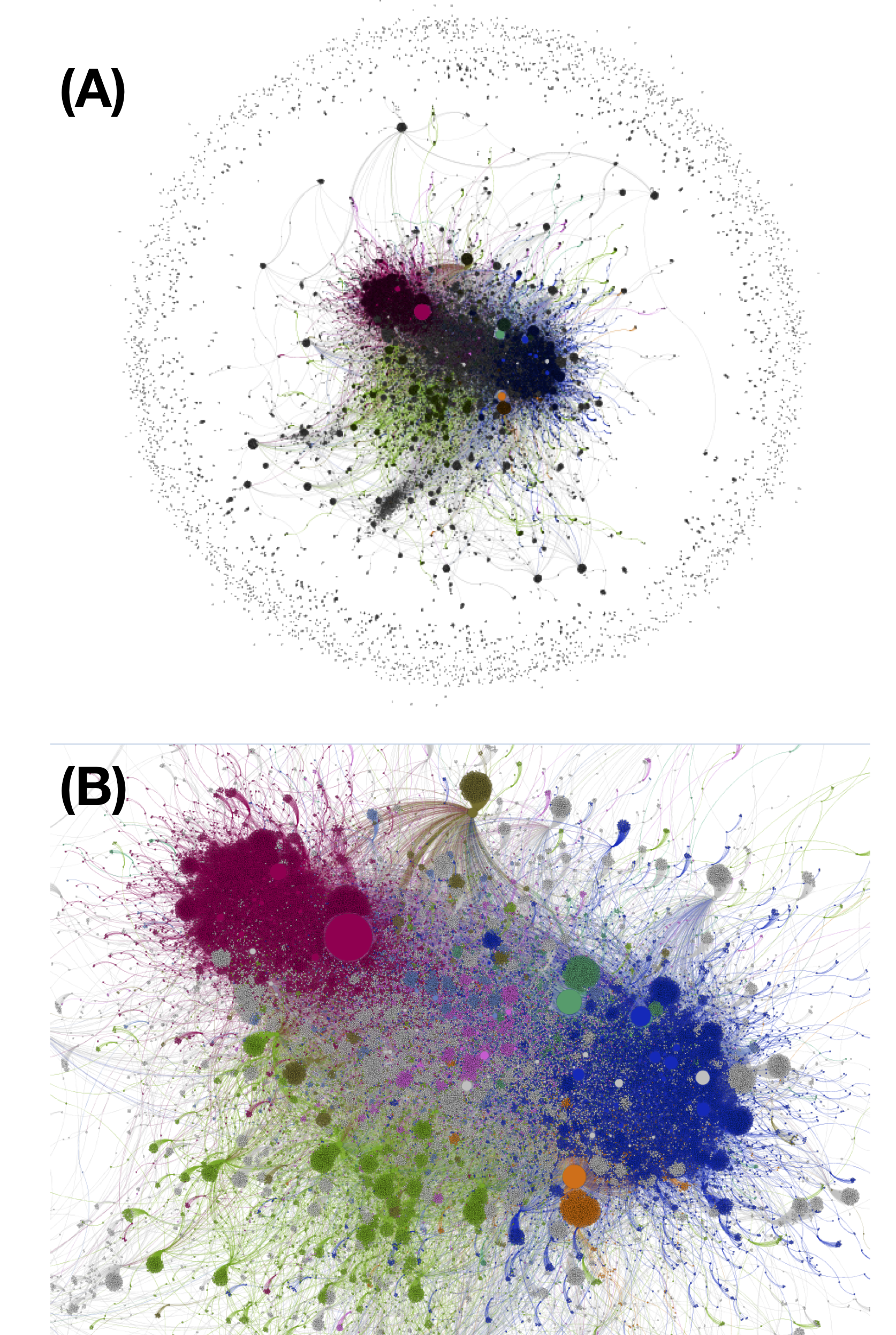}
    \caption{(A) Communities identified in political discussions shown with different colors. The communities detected using the Louvain method~\cite{Blondel_2008}. Observe that most of the nodes at the periphery have degree one, and that the network is polarized into groups with opposite points of view about the subject, in this case COVID related issues. The links are curved and if they go clockwise they are links that come out and vice versa. (B) A zoom of the network showing the two main groups.} 
    \label{fig:rt_network_communities}
\end{figure}

Fig.~\ref{fig:rt_network_communities} shows the resulting network. Different colors indicate the communities detected using the Louvain method~\cite{Blondel_2008}, which is a modularity algorithm based on optimization, by measuring the relative density of edges inside communities with respect to edges outside communities. Now two polarized groups, with opposing political points of view are clearly distinguished (pro-government in red, anti-government in blue). However, the division of these two main groups is somewhat fuzzy. We attribute it to the fact that many intermediate nodes publish topics closely related to news, notes on the advance of the pandemic and others. Tweets linking to press notes are represented by the green nodes. The little nodes represented in the exterior circumference are tweets with no impact in the discussion going on in the giant component.

\section{Analysis of the network}\label{Sec:Analysis}

\begin{figure}[htb]
    \centering
    \includegraphics[width=.95\columnwidth]{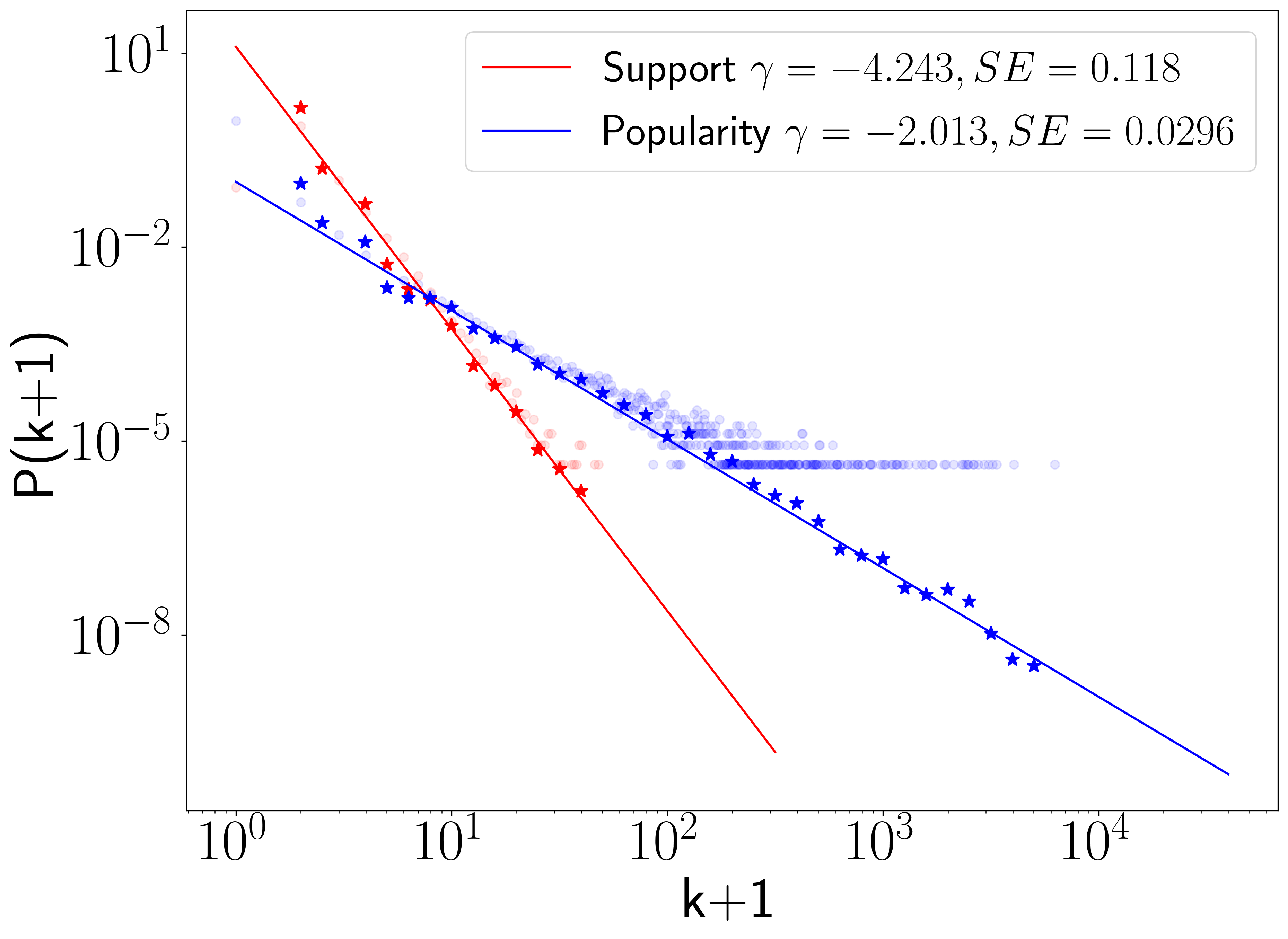}
    \caption{Log-log plot of the user popularity (blue) and number of users they support (red) distributions. The data sets in light colors correspond to raw data in a linear binning, while the points in dark colors are the results of a logarithmic binning. The solid lines correspond to power law fits, with exponents detailed in the inset.
    Observe how the support to other users decreases less sharply and is more scattered than the popularity. However, a crossover is seen near $k=10$. The displacement of $k$ by one is made to plot nodes in the logarithmic binning.}
    \label{fig:popularity_support}
\end{figure}

To obatain the parameters of the network structure, we calculated the distributions $P(k_{in}+1)$ and $P(k_{out}+1)$. Notice that here we displaced $k$ by $1$ in order to plot nodes with $k_{in}=k_{out}=0$; they form part of the network because a tweet was emitted (or received) by them although they do not have incoming or outgoing links.  $98\%$ of the nodes are of this kind. In Fig.~\ref{fig:popularity_support} we show that these distributions follow a power law behavior: $P(k_{in}) \sim (k_{in}+1)^{-2.0}$ and $P(k_{out})  \sim (k_{out}+1)^{-4.4}$, respectively.

It is important to notice in Fig.~\ref{fig:popularity_support} that most users retweet to no more than 20 different users, which implies that they are very selective when it comes to supporting an opinion. However, this is enough to generate nodes receiving thousands of retweets. Furthermore, a clear crossover is seen around $k \sim 10$, suggesting that in comparison, very popular users do not support as many users, and this behavior is reversed for the not so popular ones.
When analyzing the amount of in-degrees {\it vs.} out-degrees for the same user (data not shown), we noticed that very popular users, receiving more than 500 retweets, sent 4 or less retweets, while users with $k_{in}<=10$ sent $99.5\%$ of all retweets.

We calculated the clustering coefficient of nodes  
taking into account that the network is directed~\cite{Watts1998}:

\begin{equation}
    C_{j}=\frac{1}{k^{tot}(j)(k^{tot}(j)-1)-2k^{\leftrightarrow}(j)}A^{3}_{jj}
\end{equation}
where $\bm{A}$ is the  adjacency matrix, \emph{i.e.}, its elements are  $A_{ji}=1$ if $j$ has a link towards $i$ and zero otherwise. Notice that here we consider a directed graph and thus  $\bm{A}$ is not symmetric. In fact, $k^{tot}(j)$ is the sum of the in-degree and out-degree at node $j$,
\begin{equation}
  k^{tot}(j)=k^{in}(j)+k^{out}(j)=(\bm{A})_j \bm{1}+(\bm{A}^{T})_j \bm{1}   
\end{equation}
where $\bm{A}^{T}$ is the transpose of $\bm{A}$, $(\bm{A})_j$ stands for the $j$-th row of
$\bm{A}$, and $\bm{1}$ is the N-dimensional column vector $(1,1,...,1)^{T}$.
Also, we use the definition~\cite{Fagiolo2007}, 
 \begin{equation}
     k^{\leftrightarrow}(j)=A^{2}_{jj}
 \end{equation}
 This last term is a correction needed to avoid over-counting by revisiting loops at site $j$.

The clustering coefficient reflects the extent to which friends of $j$ are also friends of each other; and thus $C_j$ measures the cliquishness of a typical friendship circle~\cite{Watts1998}. In this particular case, it captures users that supported someone by retweeting any of their messages (or {\it vice versa}) or by retweeting each other. This in some way reflects the degree of communication between users who support a message. It is natural to calculate $C_j$ as one can suspect that the larger the network, the more difficult it would be for the members of a group to have communication with the rest of the group.

\begin{figure}[htb]
    \centering 
    
    \includegraphics[width=\columnwidth]{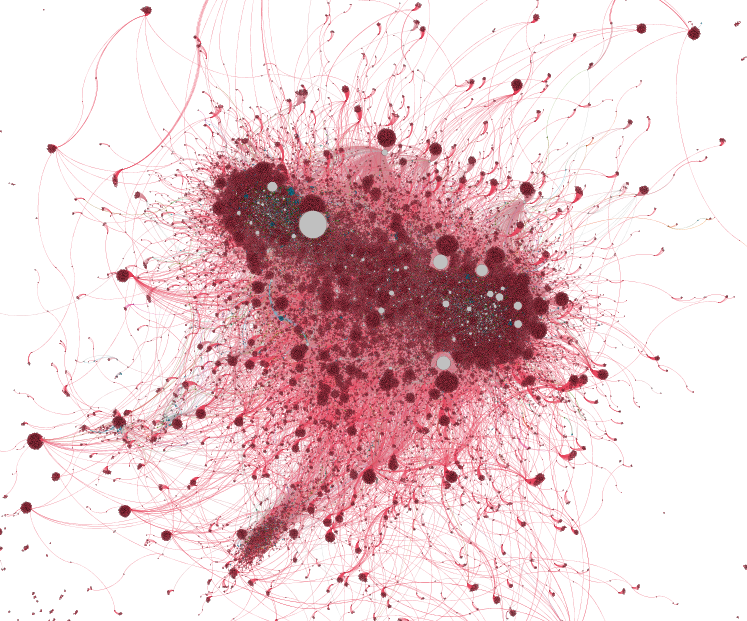}
        
    \caption{Snapshot of a retweet network over a period of approximately 3 weeks in the year $2020$, with a focus on the clustering coefficient. We associate a color with the clustering value of each node. Links are colored according to the color of the origin node. Red was assigned for clustering 0.0, blue for clustering 0.5. For this network, approximately 77\% of users have a clustering coefficient of 0.0. The size is determined by the indegree. The clockwise links are outgoing and \emph{vice versa}.}
    \label{fig:snapshot}
\end{figure}

In Fig.~\ref{fig:snapshot} we plot the retweet (Rt) network, with color based on the $C_j$, and size determined by $k_{in}$. A constant pattern is observed: if a node has many surrounding nodes, then this tends to retweet messages from one or more users, forming the well-known echo chambers~\cite{Sasahara2020}. The majority of these have a brown color, which represents $C_j=0$. Most of the members of the two main opposing groups are supported by many users (high in-degree), but they do not form clusters. Some nodes have $C_j=0.5$ (in blue) but are not easily seen since they have lower in-degrees compared with the ones already mentioned.

\begin{figure}[htb]
    \centering
    \includegraphics[width=1.0\columnwidth]{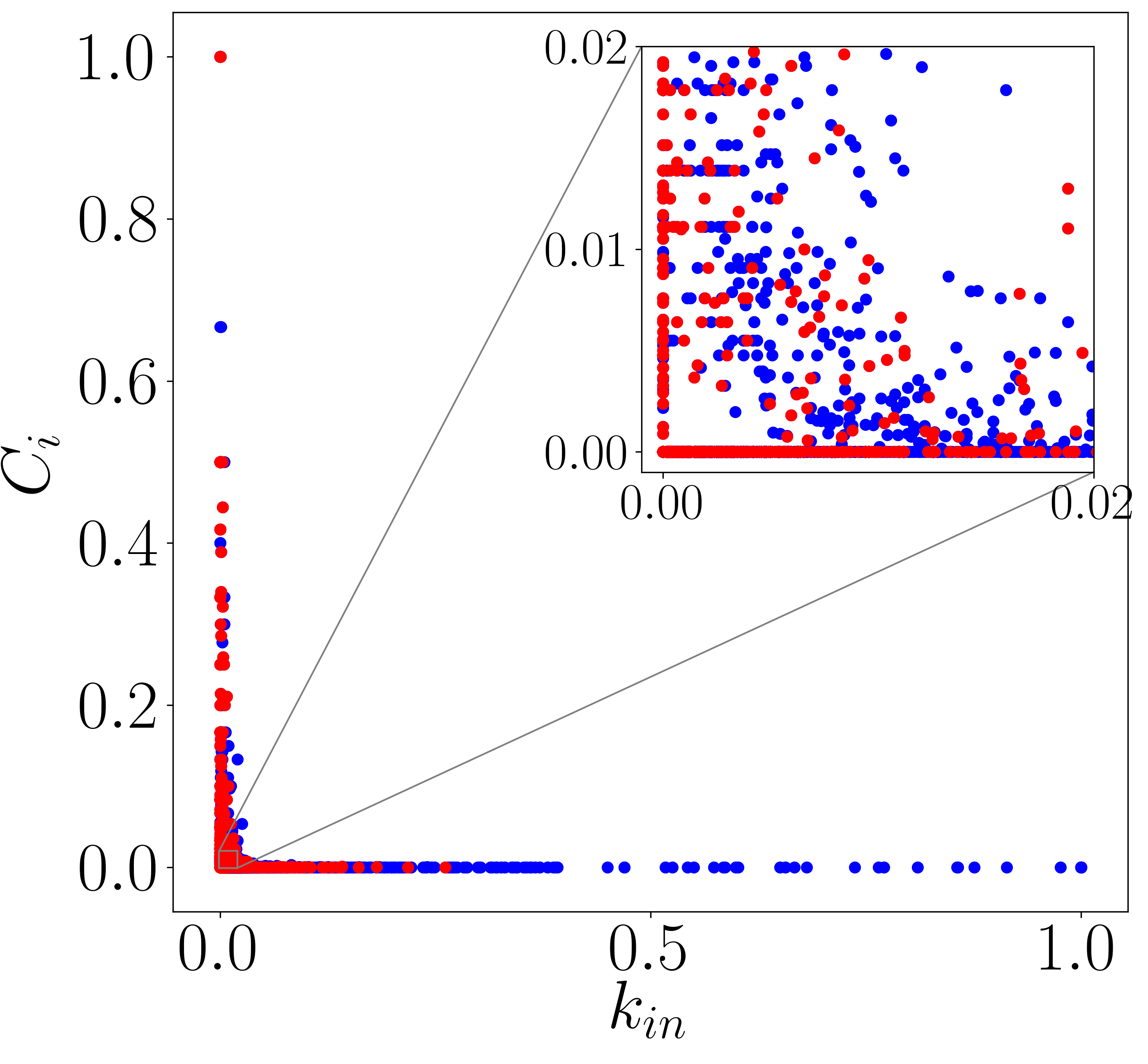}
     \caption{Clustering coefficient versus the in-degree of the node in the retweet network. We normalized $k_{in}$ and eliminated all nodes with clustering equal to zero because these nodes mostly represent the simplest dynamics in the network: users generating a retweet without some kind of prior coordination. These users represent 99.7\% of accounts. We show in red the users that we have detected have coincided at least three times by placing a hashtag or retweeting the same message with other users, these users represent 1.6\% and in blue the rest. The inset shows a zoom of the data, indicating how the network is separated into three kinds of nodes: the ones that almost fall on each axis, and a third class that does not follow such tendency and are similar to lattices obtained with dynamical feedback.}
     \label{fig:ClusteringVsDegree}
\end{figure}  

\begin{figure}[htb]
    \centering
    \includegraphics[width=.95\columnwidth]{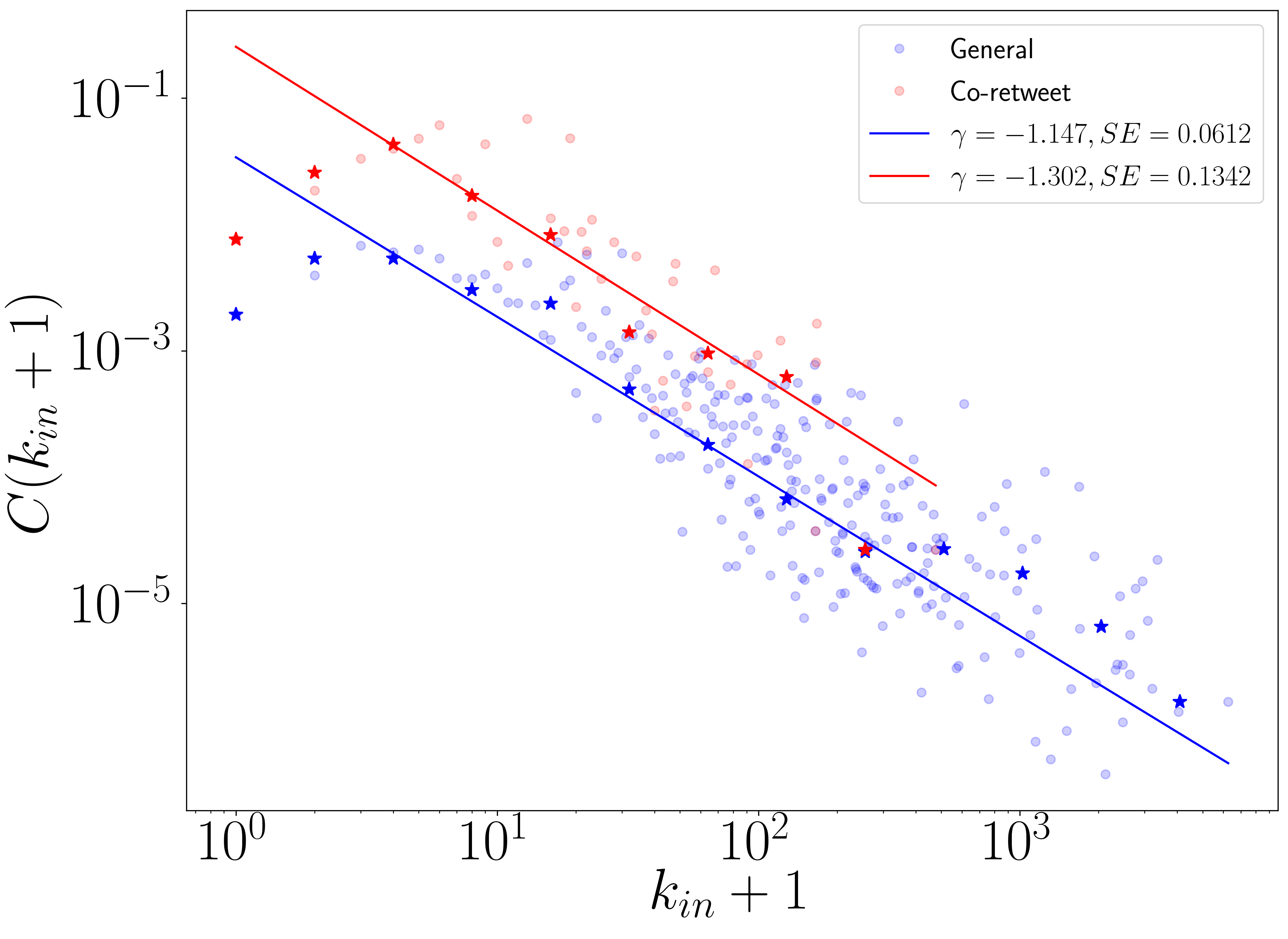}
    \caption{Log-log plot of the average clustering coefficient versus the node's in-degree $k_{in}$  seen in Fig.~\ref{fig:ClusteringVsDegree}. The raw data is presented in light colors, while data points presented in dark colors are obtained using a logarithmic binning. The power law spans nearly four decades. 
The red color denotes the subset of users which participated with some degree of organization to place a hashtag or retweet. For both sets of users a power law is detected with  exponents written in the inset, although for small in-degree, a clear departure is seen from the power law.}
    \label{fig:clustering_average}
\end{figure}

To gain a better understanding of the structure of the Rt network, we analyzed the relationship between clustering coefficient and in-degree. Fig.  ~\ref{fig:ClusteringVsDegree} shows the clustering coefficient versus the in-degree normalized to one, for data obtained from daily retweets. 
Users having $C_j>0.25$ have very low in-degrees, indicating that the clicks observed here are made up by 3 or 4 users, and even less connected groups also have few users. Users with $k_{in}>0.25$ have clustering coefficients very close to zero, supporting the observation that users receiving a large number of retweets do not have a reciprocal behavior upon other users. We identified all users coinciding (see below) at least three times in placing a hashtag or retweeting the same message during a day (red points); in general, they are indistinguishable from the rest of the users, none of them have big $k_{in}$.
From $k_{in} \sim 10$ on $C_j$ decays as a power-law with respect to $k_{in}+1$ (see Fig.~\ref{fig:clustering_average}, blue points), \emph{i.e.},
\begin{equation}
    C_j \sim \frac{C_0}{k^{\gamma}(j)}
\end{equation}
where $\gamma=1.297 \pm 13.9\mathrm{e}{-4}$ and $C_0 \pm$ 0.255 for the $k_{in}$ (popularity). Users coinciding three times or more sending the same message (red points) do have a similar behavior (same slope), but for the same values of $k_{in}$, they have higher $C_j$.

Such variation of the clustering with the degree of a network is characteristic of networks with a feedback, usually when there is a dynamical imbalance  between excitatory and inhibitory connections, as is indeed observed in many real networks~\cite{Brot2012}. In fact, the high clustering coefficient for small $k_{in}(j)$ can be explained by the transitive advantage of nodes with a common first neighbor making similar nodes to accumulate common links, and these links contribute to the similarity between the nodes~\cite{Brot2012}. Thus,  Fig.~\ref{fig:ClusteringVsDegree} hints for a more real social interaction in the network.

\section{Co-retweet Network}\label{Sec:Coretweet}
\begin{figure}[htb]
    \centering 
    \includegraphics[width=\columnwidth, height=5cm]{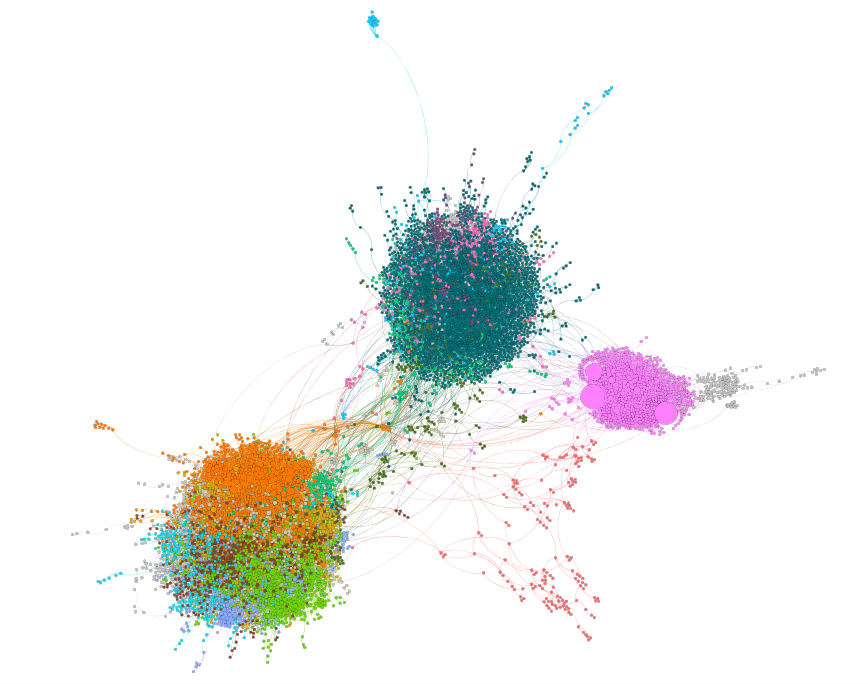}
    \caption{Co-retweet network obtained using time windows of one hour. The color was given by the community to which the user belongs. The three clusters from left to right are pro government, anti-government, and COVID--19 groups.  The size is determined by the in degree}
    \label{fig:co_retweet}
\end{figure}

Exploring the possibility of the existence of another organization level, a co-retweet network was generated~\cite{Pacheco2020Coordinated} by adding weight links between users who retweeted the same message in a given period. Weight corresponds to the number of coincidences in this time interval. The obtained network is illustrated in Fig.~\ref{fig:co_retweet}. Three groups are separated, one comprising different pro-government communities (lower right), another including anti-government communities (upper center), and a third one containing COVID--19 groups (pink nodes). The three groups form a giant component, since the network is fully connected. Links with weight $1$ were eliminated in this figure since a single coincidence may occur by chance and does not imply an underlying organization.

\begin{figure}[hbt]
    \centering
    \includegraphics[width=.8\columnwidth]{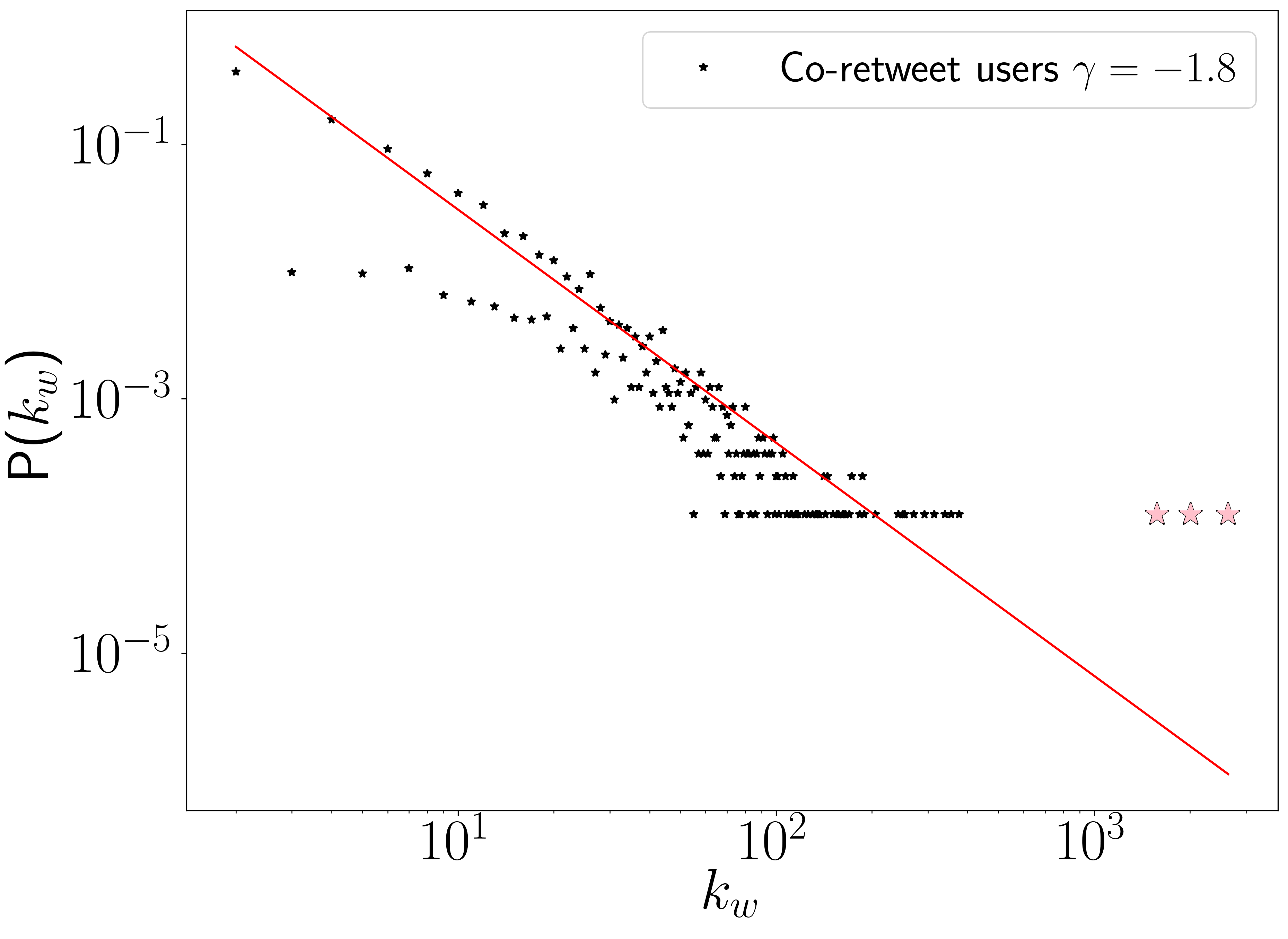}
    \caption{Degree distribution for the co-retweet network for periods of one hour. We used linear binning in this figure to highlight the outliers behavior. There are three hubs, indicated with pink stars, with a clear outlier behavior. The three hubs with highest number of retweets generate the pink-cluster community seen in Fig.~\ref{fig:co_retweet}.}
    \label{fig:outlier}
\end{figure}

The weight degree ($k_w$) distribution of this co-retweet network is shown in Fig.~\ref{fig:outlier}. Note that $3$ users have an atypical behavior. When conducting a direct search on their profiles, we found that the first one --- @CoronaUpdateBot --- is dedicated to retweeting everything it finds on coronavirus. The second  @worldnewseng retweets virtually any news. The third one is not accessible anymore @BillEsteem. All these three nodes meet the definition of superspreaders.

Trying to uncover the social network contained in the social media network, links were removed to study how the biggest component decreases as the weight of the filtered edges is progressively increased. When links weighted $1$ to $3$ were removed, the major giant component size decreased abruptly (see Fig.~\ref{fig:co_retweet}). When eliminating the links from $k_w=3$ to $k_w=48$, the size of the biggest component has a power-law decrease. In this range, the maximum giant component is much bigger than the rest of the components obtained. Filtering edges from $k_w=48$ onwards leads to a much slower decrease in the maximum component size. In this range, components obtained for each filter are more homogeneous in size. More than 48 coincidences with another user must reflect either some type of organization, or the fact that there is a social link between users.

\begin{figure}[htb]
    \centering
    \includegraphics[width=.9\columnwidth]{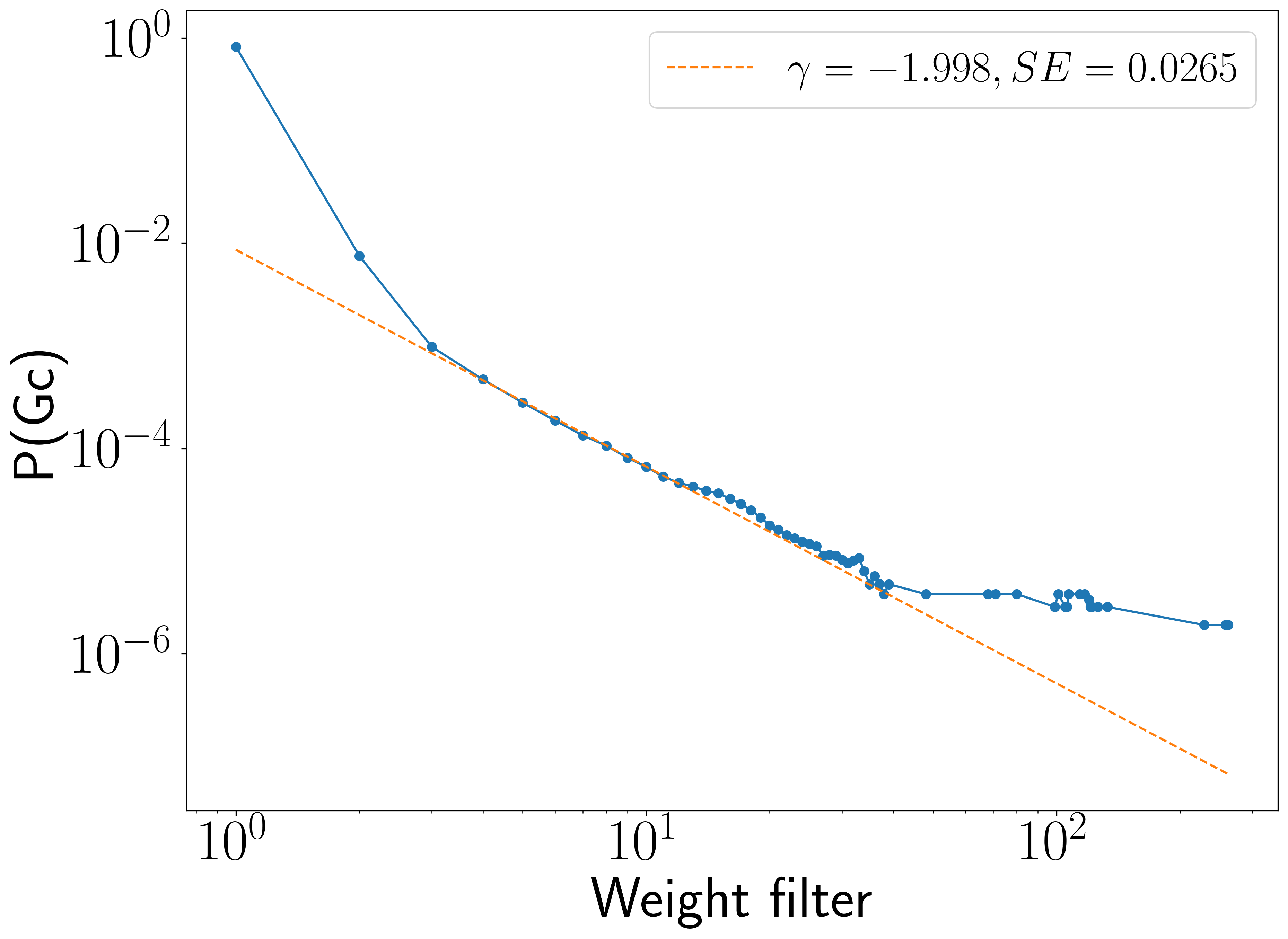}
    \caption{Giant Component size decrease as a function of the weight filter (edges being removed). The relative size of the giant component for each case is obtained by the ratio between the number of nodes in the biggest component and total number of nodes in the original co-retweet network. It is important to note that we find a region ($k_w$ between 3-40) approximately that appears to follow a power law. The co-retweet network used in this figure was obtained using windows of 1 hour.}
    \label{fig:Gc_distrib}
\end{figure}

\section{Conclusions}

Data availability has allowed only recently to study with great detail social interactions. Twitter has the peculiarity of allowing human and automatic users to participate with different purposes. Different methods have been proposed for detecting ``bots'': accounts that try to manipulate opinions through a variety of interventions. Given the impact that social networks are having on collective decision making, in issues related to elections, climate change, the COVID--19 pandemic, and many more, it is desirable to have a better understanding of the dynamics and mechanisms at play in social media. From our analysis, it can be seen that statistical methods can be useful to detect different behaviors and clearly classify most users according to them.

Summarizing, our data analysis shows that the RT network is mostly made from nodes which are very different from other social networks. In particular, we found that only  $1.4\%$ of the nodes present a clustering versus in-degree with a  power-law behavior, typical for a social network driven a feedback interaction process.  Moreover, a time-window analysis of the co-retweet network confirmed that the network is dominated by several non-human superusers.  Therefore, in this work we found a relatively simple way to separate real social interactions from other components of the lattice.

\section{Methods}\label{sec:methods}

\subsection{Data Mining Methodology}

We used the tweepy~\footnote[1]{https://www.tweepy.org/} library for data collection due to the ease of accessing the Twitter  application programming interface (API) and thus obtain the \href{https://gitlab.com/TPFBK/political-discussions-on-twitter.git}{data} in a simple way. We collected tweets for about 3 weeks of the month of May 2020 ($6{\mathrm{th}}-21^{\mathrm{st}}$). For tweet filtering, we decided not to use the filter by geolocation because it is disabled by default, and only about 1\% of the collected users have this option activated.

Subsequently, we apply a filter by pattern matching words in the text of the tweets. Once the filter is generated, we extracted the names of all the users who issued these messages, and used them as the basis for all subsequent analyzes. The filter helped us to identify the users who participated in the discussion about the COVID--19 health-emergency in Mexico. We also kept the rest of the messages to identify more general patterns of behavior. 

We analyzed the data in two different ways: For Fig.~\ref{fig:rt_network_communities}, we accumulate the connections obtained during the complete period observed. For data in Fig.~\ref{fig:ClusteringVsDegree} we generate the network of retweets for each day, in such a way that each day the network is restarted and therefore each user begins their clustering coefficient, in-degree and out-degree from zero and in the end we simply collect the measurements obtained from each day without carrying out any operation on it.

\subsection{Co-retweet network}

The co-retweet network is generated by creating a bipartite network between the user and the message that he retweets in a given period, we used periods of one hour. Given this bipartite network, we generate a projection with weights towards users, in such a way that the weight of the link gives us information on the number of matches that relate both users.

To generate the co-retweet network we must generate a two-party network, which is built from users who retweet the same message in a defined time window. So on the one hand, we have the user who makes the retweet and on the other hand, the retweeted message. We generate the link as long as they enter during the same time window. Later on, we generate a weighted projection towards the users. In this way, we obtain a weighted network that shows us in the links which users coincided in retweeting the same message in the same period of time and how many times did they coincide in it.

\section*{Acknowledgements}

We thank UNAM-DGAPA projects IN102620, IN107919, and IV100120.
RAB was financially supported by Conacyt through project 283279.

\bibliography{main.bib}

\end{document}